\documentclass[ 10pt, conference]{IEEEtran}

\usepackage[algo2e]{algorithm2e} 
\usepackage{amsmath,amssymb,amsfonts}
\usepackage{bm}
\usepackage{bbm}
\usepackage{graphicx}
\usepackage{cite}
\usepackage{float}

\usepackage{tabulary}

\usepackage{url}
\usepackage[caption=false,font=small,labelfont=sf,textfont=sf]{subfig} 
\usepackage{tabularx}
\usepackage{enumerate}
\usepackage{tikz}
\usepackage{pgfplots}
\usetikzlibrary{patterns}
\pgfplotsset{compat=newest}
\newlength\fheight
\newlength\fwidth

\usepackage{epstopdf}
\usepackage{mathtools}
\usepackage[utf8]{inputenc}
\usepackage{xcolor}
\usepackage{tikz}
\usetikzlibrary{shapes,chains,arrows,calc,positioning}
\usepackage{soul}
\usepackage{booktabs} 
\usepackage{multirow}
\usepackage{enumitem}
\usepackage{makecell}
\usepackage{lipsum}
\usepackage{graphicx}
\usepackage{pifont}
\usepackage{soul}
\sethlcolor{cyan} 
\usepackage{csquotes}

\usepackage[nolist,nohyperlinks]{acronym}



\IEEEoverridecommandlockouts
\IEEEoverridecommandlockouts\IEEEpubid{\makebox[\columnwidth]{ \hfill} \hspace{\columnsep}\makebox[\columnwidth]{ }}
\begin{document}

\title{Evaluating 5G Networks for U-Space Applications: Insights from Dense Urban Measurement Campaign}

\author{\IEEEauthorblockN{
Ricardo Barrios-Muñoz$^{\sharp}$\vspace{0.1cm},
Matteo Bernabè$^{\sharp}$, 
David L\'{o}pez-P\'{e}rez$^{\sharp}$, 
David Gomez-Barquero$^{\sharp}$, and \\
Israel Quintanilla-Garcia$^{\sharp}$}
\\ \vspace{-0.3cm}
\normalsize\IEEEauthorblockA{
\makebox[0.5\textwidth][c]{$^{\sharp}$\emph{Universitat Politècnica de València, Spain}}
\vspace{-2em}
}
\thanks{This research was supported by the Generalitat Valenciana, Spain, through the  CIDEGENT PlaGenT,  Grant CIDEXG/2022/17, Project iTENTE, and the action CNS2023-144333, financed by MCIN/AEI/10.13039/501100011033 and the European Union “NextGenerationEU”/PRTR. We also thank Rohde \& Schwarz for providing the necessary equipment.}
}


\maketitle

\begin{abstract}
Following the burgeoning interest in \acp{UAV} utilization within human-inhabited spaces, critical challenges arise in ensuring reliable, low-latency communication---particularly important given the safety-critical nature of such operations in densely populated urban environments.
Therefore, adequate cellular communication capabilities are essential to enable safe and effective operations within the so-called U-Spaces.
In this context, this paper investigates the communication performance of cellular-connected \acp{UAV} in dense urban environments. In particular, the analysis is based on a comprehensive measurement campaign conducted in the city of Benidorm, Spain---an urban area well known for its high concentration of tall buildings and overall urban density.
More specifically, we evaluated \acp{KPI} related to received signal strength and quality, data rate, and latency across various altitudes, mobile network operators, access technologies, and frequency bands, using multiple types of measurement equipment.
The results highlight significant challenges, primarily due to the lack of dedicated planning for aerial coverage and interference management, revealing that current cellular networks may fall short in supporting reliable and ubiquitous \acp{UAV} communication.
Thus, this paper calls for improved network solutions to ensure the reliability of \ac{UAV} operations in urban airspace, thereby contributing to the integration of \acp{UAV} into urban logistics and mobility.
\end{abstract}

\acresetall 
\section{Introduction}



\Acp{UAV} are transforming urban environments by enabling a range of applications, 
such as last-mile delivery, aerial surveillance, and urban mobility services like aerial taxis~\cite{ BussinesComDronesMarket2023}. 
Amazon’s Prime Air service, 
which began operations in California in 2022,
is expanding internationally to the UK and Italy, 
reflecting the growing demand for \ac{UAV}-based logistics and the role \acp{UAV} will play in urban infrastructure~\cite{AmazonPrimeAirItaly_amazon, AmazonPrimeAirItaly_times}. 



However, the integration of \acp{UAV} into urban airspace presents regulatory and safety concerns. 
Increasing \ac{UAV} traffic raises risks of conflicts with manned aircraft, airspace congestion, and noise pollution. 
These challenges underscore the need for a structured regulatory framework to ensure safe \ac{UAV} operations in urban environments\cite{doole2018drone, 9566514, 9992172, 10209247}. 
%
%
%
In response, Europe has developed the U-Space initiative~\cite{9303396}, 
designed to manage \ac{UAV} traffic in low-altitude airspace. 
This framework coordinates \ac{UAV} in populated areas, integrates automated air traffic management, and establishes safety standards, 
ensuring that \acp{UAV} can operate safely within existing air traffic systems while minimizing risks to citizens and infrastructure.



The U-space framework encompasses various services and communication requirements essential for managing drone traffic efficiently. 
The core components of U-space include the \acp{UAV}, the \ac{USP}, and the \ac{UTM} system.  
These elements must exchange real-time data to perform tasks such as flight authorization, tracking, and emergency management. 
U-space communication needs are demanding, 
with \ac{UAV} requiring reliable \ac{C2} links with latencies under 20\,ms for remote piloting,
particularly in urban areas.
The \ac{USP} and \ac{UTM} require high-throughput data exchanges (around 10\,Mbps) for monitoring \ac{UAV} positions, 
validating flight plans, and resolving conflicts.
Reliable communication is crucial for ensuring operation across these services~\cite{3GPPTS22125}.



5G, with \ac{URLLC} and \ac{mMIMO} technologies, is a strong candidate for meeting U-space’s communication needs, 
offering low latencies and high data rates crucial for \ac{UAV} command and real-time responsiveness. \cite{8918497, 10569086} 
However, a key concern remains: 
Are current cellular deployments sufficient for U-space? 
Many existing chipsets lack support for advanced 5G \ac{URLLC}, 
casting doubt on the readiness of today’s networks to reliably support U-space operations.


This paper evaluates the suitability of current commercial cellular networks for supporting U-space communication requirements. 
We present results from a measurement campaign in Benidorm, Spain, 
a key \ac{UAV} testbed location in Europe, 
known for its dense urban environment with high-rise buildings and a dense network deployment.
Moreover, it should be noted that this city has also been selected for its pivotal role in the European U-ELCOME project---an initiative involving 51 partners from Spain, France, and Italy---which enables us to conduct flights with specialized pilots in full compliance with stringent European and national regulations regarding operations in densely populated urban areas~\cite{EuropeanCommission2021, UELCOME2025}.
%
%
This evaluation aims to determine whether current network deployments are adequate or if further upgrades are necessary to meet U-space's demands; to this end, we evaluated multiple \acp{KPI} measured across various altitudes, including neighboring cell density, coverage \ac{RSRP} and \ac{SINR}, as well as \ac{UE} data throughput and latency.

The structure of the paper is as follows: 
Section \ref{sec:literature} reviews relevant literature, 
Section \ref{sec:measurement_setup} describes the measurement area and setup, 
Section \ref{sec:results} presents the findings, 
and Section \ref{sec:conclusions} discusses conclusions and potential future research directions.
\section{Literature review}
\label{sec:literature}

In the following, we present recent studies on \ac{UAV} coverage in commercial \ac{LTE} and \ac{NR} networks. Most existing research has focused on rural and suburban environments, where network conditions are relatively less complex and interference levels are moderate. In contrast, comprehensive evaluations of \acp{UAV} performance in densely populated urban areas remain limited.

The studies in~\cite{Kovacs2017-qt, Amorim2017-vk, Nwabuona2023-cw, 8301389} investigated \ac{UAV} connectivity over commercial \ac{LTE} and \ac{NR} networks in rural environments, using measurement-equipped \acp{UAV} flying at various altitudes.
Their results showed that these networks, optimised for \acp{gUE} only, caused rapid variations in performance and loss of coverage in rural areas. 
Similar results were obtained considering interference and \ac{SINR}, for which the studies in~\cite{Nwabuona2023-cw, 7470934, noauthor_undated-hl, Gharib2023-rv} highlighted a degradation of \ac{SINR} caused by increased interference stemming from the perception of more numerous network cells secondary lobes.

Limited research has addressed \ac{UAV} communication within urban environments, with existing studies primarily concentrating on low-density urban areas, thereby neglecting the characterization of highly complex and densely populated urban scenarios.
Moreover, existing studies typically rely on commercial equipment, such as mobile phones running measurement applications~\cite{Luo2022-aj, Horsmanheimo2022-tk, Sae2020-ls}. Traditional mobile phones introduce variability in results due to differences in device sensitivity and functionalities. Furthermore, these measurements are typically biased toward the performance of the specific network operator to which the phone is connected, limiting their ability to provide a generalized characterization of the surrounding environment.
In addition, since only the phone active connection is monitored, the measurements reflect the overall communication link without distinguishing between different radio access technologies, such as \ac{LTE} and \ac{NR}.
Instead, other studies have employed professional RF scanners \cite{Urama2020-xl, Nguyen2017-mt}, enabling more detailed and unbiased analyses of network behaviour without the need for a direct network connection. However, they do not simultaneously monitor overall environment coverage and specific \ac{UE} data \ac{KPI}.
Tab.~\ref{tab:uav_comparison} summarizes the key differences between these related studies and ours.

In conclusion, previous research has primarily focused on network performance in sparsely populated areas, with either RF scanners for characterizing channels or commercial phones for characterising specific \acp{KPI}.
Unlike prior work, this study evaluates the overall coverage quality across multiple network operators by leveraging advanced RF scanners and commercial smartphones to measure \acp{KPI} under realistic deployment scenarios of heterogeneous networks in dense, high-rise urban environments.

\begin{table*}[t]
    \centering
    \caption{Summary of the key differences between this work and the most closely related studies in the literature.
    }
    \label{tab:uav_comparison}
\begin{tabular}{lcccccccll}
\toprule
Ref                       &             RSRP &             SINR &            Thp DL &          Latency & LTE & NR        &  Neighboring Cells &               Environments &                Equipment \\
\midrule
\cite{Luo2022-aj}        & \text{\ding{51}} & \text{\ding{51}} & \text{\ding{55}} & \text{\ding{55}} &   \text{\ding{51}} & \text{\ding{51}} &     \text{\ding{55}} &        Suburban, Urban &              Raspberry \\
\cite{Horsmanheimo2022-tk}  & \text{\ding{51}} & \text{\ding{51}} & \text{\ding{51}} & \text{\ding{51}} &   \text{\ding{51}} & \text{\ding{51}} &     \text{\ding{55}} &                  Urban &           Nemo Handy \\
\cite{Sae2020-ls}           & \text{\ding{51}} & \text{\ding{55}} & \text{\ding{55}} & \text{\ding{55}} &   \text{\ding{51}} & \text{\ding{55}} &     \text{\ding{55}} & Urban, Suburban, Rural &         Qualipoc \& Nemo Handy \\
\cite{Urama2020-xl}         & \text{\ding{51}} & \text{\ding{55}} & \text{\ding{55}} & \text{\ding{55}} &   \text{\ding{55}} & \text{\ding{51}} &     \text{\ding{55}} &             Industrial &             R\&S TSMA6 \\
\cite{Nguyen2017-mt}        & \text{\ding{51}} & \text{\ding{51}} & \text{\ding{55}} & \text{\ding{55}} &   \text{\ding{51}} & \text{\ding{55}} &     \text{\ding{55}} &                  Rural &              R\&S TSMA6 \\
Our Study                 & \text{\ding{51}} & \text{\ding{51}} & \text{\ding{51}} & \text{\ding{51}} &   \text{\ding{51}} & \text{\ding{51}} &     \text{\ding{51}} &            Dense Urban &R\&S TSMA6B + Qualipoc \\
\bottomrule
\end{tabular}

\vspace{-1em}
\end{table*}

\section{Measurement Setup}
\label{sec:measurement_setup}
In this section, we introduce the measurement campaign area, the adopted equipment, and the monitored \acp{KPI}.

\subsection{Measurement Area}\label{subsec:MeasurementArea}

\begin{figure}[!t]
    \centering
    \includegraphics[width=\columnwidth]{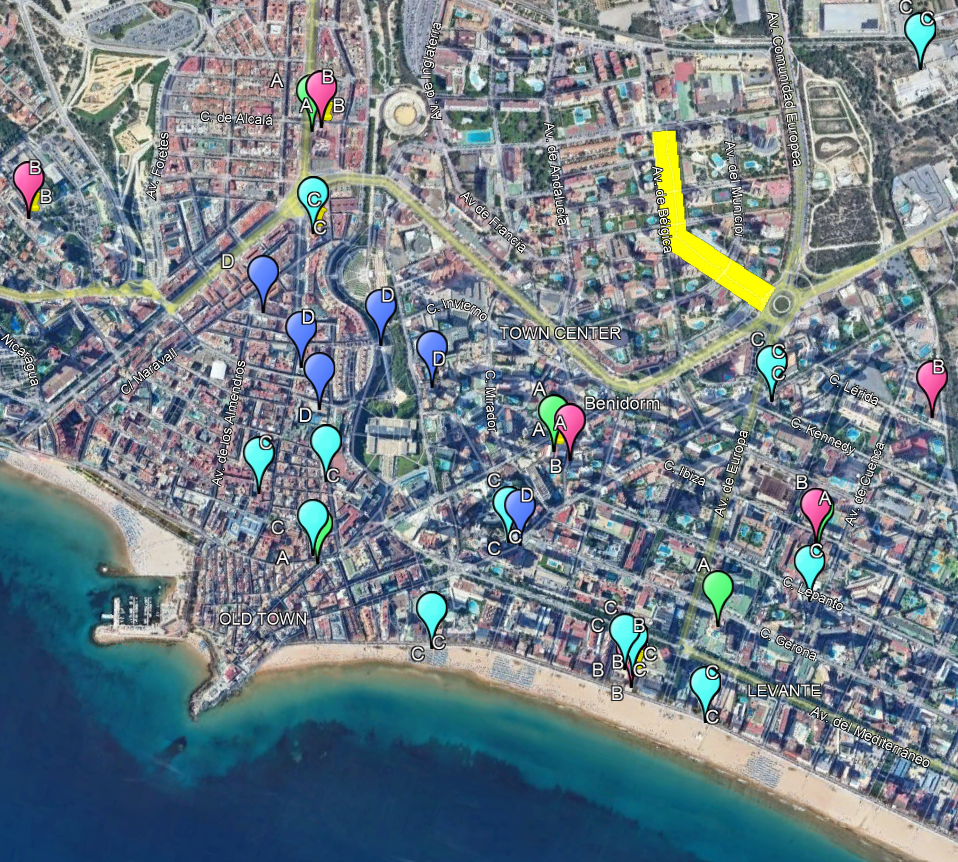}
    \caption{Benidorm, Spain, with locations of network cells from three different operators, and measurement area highlighted in yellow (top right).}
    \label{fig:AreaBenidorm}
\end{figure}

The measurement campaign was conducted in Benidorm, Spain. Often referred to as the \enquote{New York of the Mediterranean}, the city is characterized by its high population density and a remarkable concentration of tall buildings\footnote{The average building height is estimated to be slightly above 100\,m.}, which makes \acp{UAV} operations challenging in terms of both safe navigation and reliable communication.
The overall area hosts both 4G \ac{LTE} and 5G \ac{NR} networks from different operators, such as Telefónica, Orange, Vodafone, and Xfera Móviles, all operating in various frequency bands to meet urban coverage demands.
Fig.~\ref{fig:AreaBenidorm} illustrates the city of Benidorm, the locations of the measured cells from multiple operators, and the considered measured area. 
This area was chosen primarily for its dense concentration of tall buildings, which provides unique propagation conditions absent in rural and suburban settings.

The aerial measurements were conducted at three distinct altitudes: 20, 40, and 60\,m. At each altitude, the \ac{UAV} maintained a constant speed of 4\,m/s for a total duration of 15\,min, allowing for precise spatial analysis across the entire measurement area divided into a dense grid.\footnote{It should be noted that the flight duration was constrained by the \ac{UAV} payload capacity, i.e., the equipment described in Section~\ref{subsec:Measurement Equipment}.}
Then, in order to provide ground \acp{KPI} reference, later referred to as \ac{gUE}, measurements were conducted at 1.5\,m height around the reference area.

It should be emphasized that, due to the area's dense population and the sensitive nature of the operation, all the necessary permissions were obtained from the local police and relevant national authorities to ensure full compliance with regulations, including restrictions on maximum flight altitude and requirements for operating in densely populated areas from certified pilots.

\begin{table*}[t]
\centering
\caption{TSMA6B scanner measured frequency bands.}
\vspace{-0.2cm}
\begin{tabular}{|c c c c c c c c c c c|}
\hline
\textbf{LTE} & B20 & B28 & B8 & B3 & - & B1 & - & B38 & B7 & - \\ 
\textbf{NR}  & n20 & n28 & n8 & n3 & n39 & n1 & n40 & n38 & n7 & n78 \\ 
\textbf{Range (MHz)} 
& \multicolumn{2}{c}{758-821} 
& 925-960 
& 1805-1880
& 1880-1920 
& 2110-2170 
& 2300-2400 
& 2570-2620 
& 2620-2690 
& 3300-3800 \\ \hline
\end{tabular}
\label{tab:ScannerFrequency_Bands}
\end{table*}

\subsection{Measurement Equipment}\label{subsec:Measurement Equipment}

In the following, we introduce the equipment used in this measurement campaign.

\subsubsection*{UAV}
\begin{figure}[!t]
\centering
\includegraphics[width=0.45\textwidth]{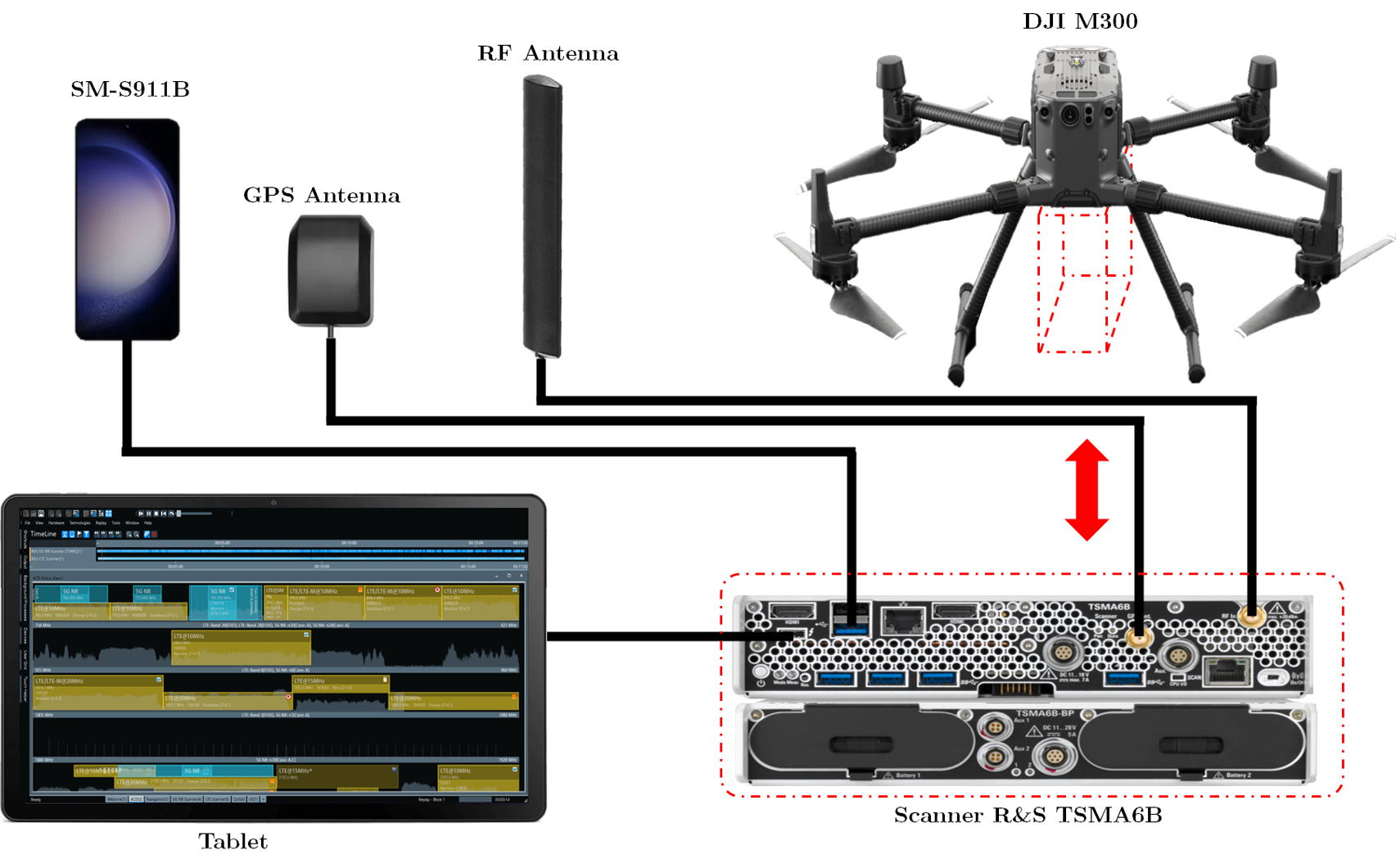}
        \caption{Measurement setup with an R\&S QualiPoc engineering phone, GPS and antenna linked to an R\&S TSMA6B scanner, mounted on a DJI M300.}
        \label{fig:equipos}
\end{figure}

The \ac{UAV} selected for the measurements was a commercial DJI M300,
chosen for its high payload capacity of approximately 3\,kg and its stable flight
---both critical for carrying the measurement equipment. 
A custom mounting platform was designed to securely hold the R\&S TSMA6B scanner, a Samsung SM-S911B smartphone, an RF antenna, and a GPS antenna 
(see Fig.~\ref{fig:equipos}). 
To optimize weight and extend flight time,
the TSMA6B scanner was operated using a single battery, 
resulting in a total payload weight of approximately 2.7\,kg across all components.

\subsubsection*{Scanner}
To accurately evaluate coverage signalling within the surrounding public networks, the Rohde\&Schwarz (R\&S) TSMA6B scanner was utilized due to its state-of-the-art capabilities.
The R\&S TSMA6B scanner was configured in passive mode to measure downlink reference signals across all frequency bands listed in Tab.~\ref{tab:ScannerFrequency_Bands}, scanning up to 32 cells for band for both \ac{LTE} and \ac{NR}.
It should be noted that the scanner can measure reference signals from all found cells without establishing a connection with any of them, therefore providing a large overview of coverage signalling from different network operators.

\subsubsection*{Engineering Phone}
To analyze performance from a typical \ac{UE} perspective, we conduct measurements using a Samsung SM-S911B phone equipped with R\&S QualiPoc software, a specialized tool designed to collect and analyze network performance data and \ac{QoS} metrics. 
Unlike the TSMA6B scanner, which passively scans and reports measurements across multiple cells, the SM-S911B integrate a SIM card and connects to a single serving cell per time. This setup enables real-time monitoring of session-level \acp{KPI}.
It is important to emphasize that the measurements obtained via the SM-S911B phone offer a more representative view of a conventional user experience. This is because the TSMA6B scanner features a higher-grade receiver, which may yield optimistically skewed performance metrics compared to those perceived by standard consumer devices.

\subsubsection*{\ac{QoS} Measured \acp{KPI} } \label{subsubsec:QoSMeasuredKPI}
To evaluate coverage \ac{QoS}, we first assess the number of different cells, each characterized by a different \ac{PCI}, detected at various altitudes. 
Then,
we focus on the measured \acp{SS} signals \ac{RSRP} and \ac{SINR} from all discovered 4G/5G cells by the R\&S TSMA6B scanner, along with those obtained by the SM-S911B phone.
Additionally, to study throughput and \ac{RTT}, iPerf3 tests were conducted with the QualiPoc phone.
Specifically, an iPerf3 server was configured in the university lab to support two concurrent TCP sessions for the tested phone, with each session dedicated to uplink and downlink traffic, respectively. Then, each was assigned a target bandwidth of 500\,Mbit/s, a transfer duration of 30\,s, and a maximum test duration of 60\,s; then, during each measurement, the devices transmitted packets with a size of 2048\,bytes.
\section{Measurement Result and Discussion}
\label{sec:results}

This section discusses the results obtained in our measurement campaign over the area described in Section~\ref{subsec:MeasurementArea}.
Specifically, data are gathered from 4 different major operators in Spain, here referred to as A, B, C and D, for which their distribution of cell is depicted in Fig.~\ref{fig:AreaBenidorm}.
Then, as outlined in Section~\ref{sec:measurement_setup}, 
the TSMA6B scanner's high sensitivity enables capturing up to 32 cells per frequency. 
Here, we focus on analyzing the Top 1---the highest ranked---within each and 
for the best operator serving the area.

Without loss of generality and for the sake of space, we only present the analysis for the following frequency bands and technologies: B3 for \ac{LTE} and n78 for \ac{NR}, where, for each, we present the main \acp{KPI} presented in Section~\ref{subsubsec:QoSMeasuredKPI}.
Similar results and conclusions also apply to all frequency bands reported by the R\&S TSMA6B scanner.

\subsection{Number of Neighbors Cells Analysis} 
\label{subsec:NumberOfNeighbors}
Fig.~\ref{fig:PCI_Number_Detected} illustrates how the total number of scanned \acp{PCI} (i.e., different cells) across all frequency bands and technologies varies with altitude during the \ac{UAV} flight.
A clear trend is observed: as the altitude increases, so does the number of neighbouring cells detected. For n78 (\ac{NR}), this number rises from approximately 33 to 43 \ac{PCI}'s between 20\,m and 60\,m, with a similar pattern found in B3 (\ac{LTE}).

This effect is primarily due to improved \ac{LoS} conditions at higher altitudes, allowing the UAV to \enquote{see} more cells and benefit from better propagation paths. These findings are essential for evaluating \ac{UAV} performance, as higher cell visibility may lead to increased interference and a greater likelihood of handover occurrences.

\begin{figure}[!t]
    \centering                                   
    \subfloat {
        \setlength\fwidth{0.8\columnwidth}
        \setlength\fheight{0.6026\columnwidth}
        \begin{tikzpicture}

\definecolor{darkred}{RGB}{139,0,0}
\definecolor{darkblue}{RGB}{0,0,139}
\definecolor{darkgreen}{RGB}{34,139,34} 
\definecolor{darkgray176}{RGB}{176,176,176}
\definecolor{lightgray204}{RGB}{204,204,204}

\begin{axis}[
    width=\fwidth,
    height=\fheight,
    fill opacity=1,
    legend style={
        at={(0.05,0.95)},
        anchor=north west,
        legend columns=1,
        fill opacity=0.9,
        font=\small,
        draw=darkgray176
    },
    bar width=12pt, 
    ybar,
    enlarge x limits=0.75, 
    ymin=0, ymax=49,
    ylabel={\#\acp{PCI}},
    symbolic x coords={B3-1800, 3500-n78},
    xlabel={Frequency Band [MHz]},
    xtick=data,
    xticklabels={
        \makecell{B3} (LTE),
        \makecell{n78} (NR)
    },
    nodes near coords,
    nodes near coords align={vertical},
    nodes near coords style={font=\footnotesize}, 
    x tick label style={font=\footnotesize},
    y tick label style={font=\footnotesize},
    ytick={0, 10, ..., 47},
    xmajorgrids,
    x grid style={darkgray176, opacity=0.75},
    ymajorgrids,
    y grid style={darkgray176, opacity=0.75},
    tick align=outside,
    tick pos=left,
    xtick style={color=black},
    ytick style={color=black},
    xlabel style={font=\footnotesize},
    ylabel style={font=\footnotesize},
    yticklabel style={font=\footnotesize},
    xticklabel style={font=\footnotesize},
]

\addplot[fill=darkblue, fill opacity=0.75] coordinates {(B3-1800, 20) (3500-n78, 33)};
\addlegendentry{20m}

\addplot[fill=darkred, fill opacity=0.75] coordinates {(B3-1800, 30) (3500-n78, 38)};
\addlegendentry{40m}

\addplot[fill=darkgreen, fill opacity=0.75] coordinates {(B3-1800, 40) (3500-n78, 43)};
\addlegendentry{60m}

\end{axis}

\end{tikzpicture}
    }
    \vspace{-1em}
    \caption{Number of \acp{PCI} detected at different altitudes.\label{fig:PCI_Number_Detected}}
    \vspace{-1.5em}
\end{figure}

\subsection{TSMA6B Scanner RSRP Analysis}
\label{subsec:MeasuredRSRP}

\begin{figure*}[!t]
    \subfloat [\footnotesize{RSRP TSMA6B Scanner}\label{subfig:RSRP__LTE_NR_Scanner}]{
        \setlength\fwidth{0.975\columnwidth}
        \setlength\fheight{0.6026\columnwidth}
        \input{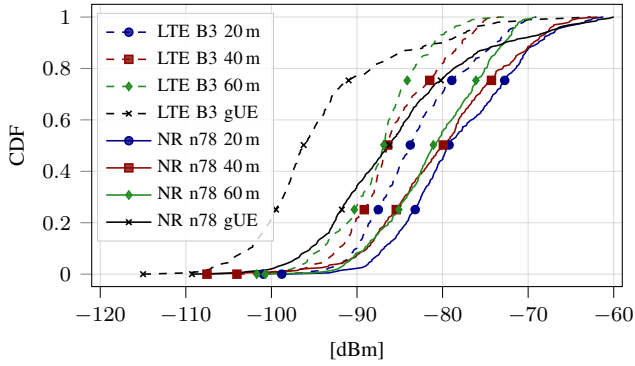}
    }
    \subfloat[\footnotesize{SINR TSMA6B Scanner}\label{subfig:SINR__LTE_NR__Scanner}]{
        \setlength\fwidth{0.975\columnwidth}
        \setlength\fheight{0.6026\columnwidth}
        \input{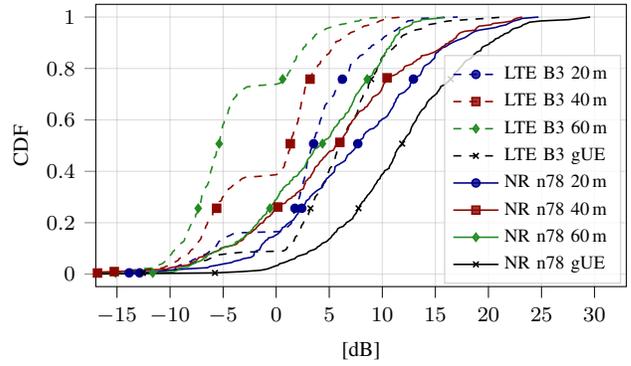}
    }

    \caption{\Acp{CDF} of \ac{RSRP} and \ac{SINR} measurements from the TSMA6B scanner at different altitudes for \ac{LTE} B3 and \ac{NR} n78 frequency bands.\label{fig:RSRP_SINR__LTE_NR__Scanner}}
\end{figure*}



In the following, we present and discuss results obtained for measured \ac{RSRP} by the TSMA6B scanner over the considered frequency bands and altitudes.
Fig.~\ref{subfig:RSRP__LTE_NR_Scanner} shows the obtained \acp{RSRP} \acp{CDF} and 
Tab.~\ref{tab:SummaryResultsStats} summarize their key values.

By analyzing the results, we can observe that
n78 (\ac{NR}) outperforms B3 (\ac{LTE}) in terms of \ac{RSRP}, despite operating at a higher frequency where propagation conditional are typically poorer.
In particular, at 20\,m, \ac{NR} achieves a mean value of -78.2\,dBm, which is 4.8\,dB stronger than that of \ac{LTE}, whose mean value reaches -83\,dBm.
These improvements can be primarily attributed to the denser deployment of \ac{NR} and its advanced beamforming capabilities.
Then, it has to be noted that the \ac{UAV} consistently outperforms the \ac{gUE} in terms of \ac{RSRP} at all altitudes, primarily due to a higher probability of \ac{LoS} conditions. In n78 (\ac{NR}), the mean \ac{RSRP} improves from –85.3\,dBm at ground level (\ac{gUE}) to –78.2\,dBm at 20\,m altitude, representing a gain of 7.1\,dB. A similar trend is observed in B3 (\ac{LTE}), where the gain reaches 11.2\,dB, with \ac{RSRP} passing from –94.2\,dBm (\ac{gUE}) to –83.0\,dBm at 20\,m.
Finally, we observe that the \ac{RSRP} tends to decrease while increasing the altitude, losing 2.7\,dB for n78 (\ac{NR}) and 4.2\,dB for B3 (\ac{LTE}) when passing from 20\,m to 60\,m.
This observed decline is attributed mainly to the diminished beamforming gain, as the \ac{UAV} deviates from the downtilted main lobe direction optimized for ground coverage.

These results demonstrate that \acp{UAV} operating in urban environments maintain consistently high \ac{RSRP} levels. Given the conventional threshold of -119\,dBm, the observed values surpass this by approximately 30\,dB, confirming that received signal power is not a limiting factor in such scenarios.

\subsection{TSMA6B Scanner SINR Analysis}
    
In the following, we present and discuss the results obtained for the measured \ac{SINR} by the TSMA6B scanner over the considered frequency bands and altitudes.
Fig.~\ref{subfig:SINR__LTE_NR__Scanner} shows the obtained \acp{RSRP} \acp{CDF} and 
Tab.~\ref{tab:SummaryResultsStats} reports values.

Observing the obtained results,
we can identify the following multiple key aspects.
First, although \ac{UAV} register stronger \ac{RSRP} values, \ac{gUE} consistently achieve better \ac{SINR} for both n78 (\ac{NR}) and B3 (\ac{LTE}). 
This is due to the likely \ac{NLoS} conditions at ground level, where urban obstructions act as natural barriers, significantly attenuating interference.
Then, we note that
\ac{SINR} performance deteriorates with increasing altitude. This is caused by both decreased \ac{RSRP} and stronger interference, as higher \ac{LoS} probability makes stronger interferers more likely to be received.
Finally, considering an out-of-service threshold of -6\,dB, we observe that the potential outage rate increases with altitude. In particular, n78 (\ac{NR}) shows outage rates of 3.14\%, 8.36\%, and 8.39\% at 20, 40, and 60\,m, respectively, while B3 (\ac{LTE}) exhibits, for same altitudes, outage rates of 9.81\%, 22.92\%, and 41.81\%.

These results underscore that neither \ac{LTE} nor \ac{NR} can fully ensure reliable coverage, presenting a considerable challenge for the deployment of U-space services.
Nevertheless, it is noteworthy that n78 (\ac{NR}) exhibits significantly lower outage rates, suggesting that it may represent the most promising technology in this context.

\subsection{Qualipoc Phone Rate and RTT Analysis}

Here, we present throughput and \ac{RTT} results following the methodology described in Section~\ref{subsubsec:QoSMeasuredKPI}. Results show a degradation in downlink throughput and latency with increasing altitude. Specifically, the median downlink throughput decreases from 105.65\,Mbps at ground level (\ac{gUE}) to 41.45\,Mbps at 40\,m.
Then, the \ac{RTT} follow a similar trend, with values passing from 54\,ms (\ac{gUE}) to nearly 60\,ms at higher altitudes.

Overall, the results highlight the potential network limitation for \acp{UAV} operations, which require reliable, high-throughput and low-latency links, highlighting dependency on the height and underscoring the need for further dedicated system design and optimized network algorithms.

The obtained results are summarized in Tab.~\ref{tab:SummaryResultsStats}.
\begin{table}[!t]
\centering
\begin{tabular}{l l c c c c}
 &   & \textbf{gUE} & \textbf{20\,m} & \textbf{40\,m} & \textbf{60\,m} \\ \hline

\multirow{8}{*}{\makecell{\textbf{RSRP} [dBm] }} 
& & \multicolumn{4}{c}{\textbf{TSMA6B B3 (LTE)}} \\
& 5\%-Tile & -104.9 & -91.5 & -93.7 & -95.8 \\
& Median & -96.3 & -83.8 & -86.4 & -86.9 \\
& Mean & -94.2 & -83 & -85.7 & -87.2 \\

& & \multicolumn{4}{c}{\textbf{TSMA6B n78 (NR)}} \\
& 5\%-Tile & -97.6 & -88.3 & -91.4 & -90.9 \\
& Median & -86.2 & -79.3 & -80 & -81.1 \\
& Mean & -85.3 & -78.2 & -79.9 & -80.9 \\ \hline

\multirow{8}{*}{\textbf{SINR} [dB]} 
& & \multicolumn{4}{c}{\textbf{TSMA6B B3 (LTE) }} \\ 
& 5\%-Tile & -7.1 & -8 & -9.4 & -10.5 \\
& Median & 6 & 3.4 & 1.3 & -5.6 \\
& Mean & 5.6 & 3 & -0.6 & -4.2 \\
& & \multicolumn{4}{c}{\textbf{TSMA6B n78 (NR)}} \\
& 5\%-Tile & 1.1 & -4 & -7.6 & -7.2 \\
& Median & 11.7 & 7.5 & 5.7 & 4.1 \\
& Mean & 11.8 & 7.5 & 5.4 & 3.6 \\ \hline

& & \multicolumn{4}{c}{\textbf{Qualipoc Phone}} \\
\multirow{3}{*}{\textbf{Thp DL} [Mbps]} 
& 5\%-Tile & 10.80 & 10.96 & 9.21 & 9.71 \\
& Median & 105.65 & 99.60 & 41.45 & 66.0 \\
& Mean & 110.18 & 140.65 & 88.73 & 95.34 \\ \hline

\multirow{3}{*}{\textbf{RTT [ms]}} 
& 5\%-Tile & 42.74 & 53.57 & 54.50 & 53.42 \\
& Median & 54.0 & 55.90 & 58.03 & 59.91 \\
& Mean & 62.09 & 58.65 & 59.98 & 66.41 \\ \hline

\end{tabular}
\vspace{0.5em}
\caption{Summary of the obtained statistics for TSMA6B scanner \ac{RSRP}, \ac{SINR} and Qualipoc phone throughput and RTT.\label{tab:SummaryResultsStats}}
\vspace{-1em}
\end{table}

\subsection{Impact of Network Deployment}

In the following, 
we present and discuss the result obtained for the four different networks serving the study area,
within B3 (\ac{LTE}) and n78 (\ac{NR}) networks. 
Referred here as Network A, B, C and D for the sake of confidentiality. 
It should be noted that although these networks operate within the same frequency band, 
they do not share additional characteristics, 
having distinct and independent network deployments and configurations.

\begin{figure}[!t]
    \centering
    \subfloat {
        \setlength\fwidth{0.975\columnwidth}
        \setlength\fheight{0.6026\columnwidth}
        \input{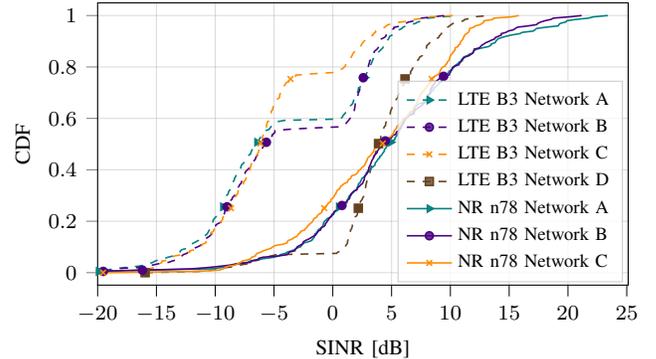}
    }
    \vspace{-1em}
    \caption{\ac{SINR} \ac{CDF} captured at 60\,m height for different network operating within \ac{LTE} B3 and \ac{NR} n78 frequency band.}\label{fig:DifferentNetworkOperator}
    \vspace{-1em}
\end{figure}

Fig.~\ref{fig:DifferentNetworkOperator} show the obtained \ac{CDF} for the measured \ac{SINR} within the \acp{UAV} flying at an altitude of 60\,m.
Analyzing the obtained results, 
we can observe how the different network deployments and configurations can significantly influence the \acp{UAV} performance, 
with median values differing by approximately 10\,dB when considering \ac{LTE}.
Therefore, these results underscore the importance of optimal deployments and configurations to support U-space operations, 
highlighting the need for effective network planning and optimization to reshape current networks or guide future 6G ones.

\subsection{Impact of \ac{UE}'s Receiver}

\begin{figure}[!t]
    \centering
    \subfloat {
        \setlength\fwidth{0.975\columnwidth}
        \setlength\fheight{0.6026\columnwidth}
        \input{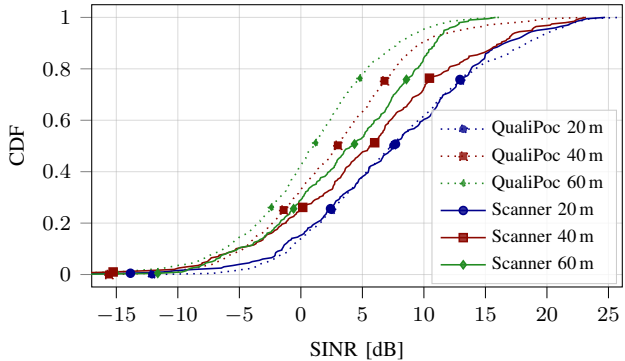}
    }
    \vspace{-1em}
    \caption{Samsung SM-S911B R\&S QualiPoc \ac{SINR} \acp{CDF} at multiple altitudes for \ac{LTE} B3 and \ac{NR} n78 frequency bands.\label{fig:Qualipoc_RSRP_SINR_NR}}
\end{figure}

The quality of the receiver has an impact on \acp{UE} performance too. 
To analyze this effect, 
we present measurements obtained by the Samsung SM-S911B phone, 
equipped with R\&S QualiPoc software, 
in terms of \ac{SINR}. 
For comparison, 
Fig.~\ref{fig:Qualipoc_RSRP_SINR_NR} shows the \ac{SINR} \acp{CDF} from the SM-S911B phone and the TSMA6B scanner, 
with both devices measuring the same operator, frequency band, and technology.

Our results indicate that the phone consistently reports lower \acp{SINR} than the scanner,
with values deviating further as the flight altitude increases.
This discrepancy likely arises due to receiver's quality. 
Moreover, it should be noted that, 
unlike the scanner, 
the phone is not always connected to the strongest cell due to the nature of handover process. 
As the \ac{UAV} moves, 
the phone may temporarily connect to weaker cells, 
resulting in lower \ac{SINR} readings compared to the scanner, 
which consistently monitors the strongest signals.

Overall, this result highlights a crucial point: 
While high-end scanners, like the TSMA6B, can be considered as high-performance \ac{UE} 
(representing an idealized \ac{UAV} receiver), 
the actual performance of \acp{UAV} in U-space will depend on the quality of the receiver used.
As the quality of the receiver improves,
so does the signal quality, 
but this comes at a higher cost. 
Therefore, \acp{UAV} equipped with better receivers can expect improved signal reception, 
but this needs to be weighed against the increased price and practical deployment considerations.

\section{Conclusion}
\label{sec:conclusions}

In this study, we evaluated coverage and communication performance of real \acp{UAV} in the dense urban environment of Benidorm, Spain, characterized by high-rise buildings and dense networks.

We analyzed different \acp{KPI} across various network technologies, frequencies, and equipment. Our results highlighted that current networks are insufficient to provide reliable and stable connectivity in the sky, which is crucial for enabling safe and secure \acp{UAV} operations.

Specifically, due to the unique \ac{LoS} conditions, despite high \acp{RSRP}, \acp{UAV} experienced poor \acp{SINR}. This resulted in high out-of-coverage rates, with values up to almost 41\% for \ac{LTE} and 10\% for \ac{NR}.

Finally, our findings hint the need for new optimization frameworks to optimally re-shape current networks or optimally plan future 6G ones for U-space safe operations.

\begin{acronym}[AAAAAAAAA]
    \acro{2D-DFT}{two dimensional discrete Fourier transform}
    \acro{3GPP}{3rd Generation Partnership Project}
    \acro{4G}{fourth generation}
    \acro{5G}{fifth generation}
    \acro{6G}{sixth generation}
    \acro{AF}{array factor}
    \acro{AH}{aerial highway}
    \acro{AI}{artificial intelligence}
    \acro{AoA}{angle of arrival}
    \acro{AoD}{angle of departure}
    \acro{BBU}{baseband unit}
    \acro{BO}{bayesian optimization}
    \acro{BS}{base station}
    \acro{BVLoS}{beyond visual line of sight}
    \acro{CAGR}{compound annual growth rate}
    \acro{CCUAV}{cellular connected unmanned aerial vehicle}
    \acro{CDF}{cumulative distribution function}
    \acro{cm-wave}{centimeter wave}
    \acro{CQI}{channel quality indicator}
    \acro{CRS}{common reference signal}
    \acro{CSI}{channel state information}
    \acro{CSI-RS}{channel state information-reference signal}
    \acro{D2D}{device to device}
    \acro{DFT}{discrete Fourier transform}
    \acro{DL}{downlink}
    \acro{DoF}{degree of freedom}
    \acro{eGA}{elite genetic algorithm}
    \acro{eICIC}{enhanced inter-cell interference coordination}
    \acro{E}{eastern}
    \acro{ES}{Eigenscore}
    \acro{FR1}{frequency range 1}
    \acro{FR2}{frequency range 2}
    \acro{FR3}{frequency range 3}
    \acro{KPI}{key performance indicator}
    \acro{USP}{U-space Service Provider}
    \acro{UTM}{Unmanned Traffic Management}
    \acro{C2}{command and control}
    \acro{GA}{genetic algorithm}
    \acro{gUE}{ground user equipment}
    \acro{HO}{handover}
    \acro{ICC}{international conference on communications}
    \acro{IMT}{international mobile telecommunication system}
    \acro{ISD}{inter-site distance}
    \acro{IUD}{inter-UAV distance}
    \acro{ITU}{international telecommunication union}
    \acro{LEO}{low Earth orbit}
    \acro{LoS}{Line-of-Sight}
    \acro{LTE}{long term evolution}
    \acro{MAMA}{mMIMO-Aerial-Metric-Association}
    \acro{MCPA}{multicarrier power amplifier}
    \acro{MINP}{mixed-integer nonlinear problem}
    \acro{ML}{machine learning}
    \acro{MIMO}{multiple-input multiple-output}
    \acro{mMIMO}{massive multiple-input multiple-output}
    \acro{mmWave}{millimeter wave}
    \acro{MNO}{mobile network operator}
    \acro{MU-mMIMO}{multi-user massive multiple-input multiple-output}
    \acro{MU-MIMO}{multi-user multiple-input multiple-output}
    \acro{NLoS}{Non-Line-of-Sight}
    \acro{NOMA}{non-orthogonal multiple access}
    \acro{NR}{new radio}
    \acro{PBCH}{physical broadcast channel}
    \acro{P2P}{point to point}
    \acro{PAHSS}{Particle Aerial Highway Swarm Segmentation}
    \acro{PL}{path loss}
    \acro{PMI}{precoding matrix indicator}
    \acro{PRB}{physical resource block}
    \acro{PSO}{particle swarm optimization}
    \acro{PSS}{primary synchronization signal}
    \acro{QoS}{Quality of Services}
    \acro{RAN}{radio access network}
    \acro{RE}{resource element}
    \acro{RI}{rank indicator}
    \acro{RRC}{radio resource control}
    \acro{RSRP}{reference signal received power}
    \acro{RSRQ}{reference signal received quality}
    \acro{RSS}{received signal strength}
    \acro{RSSI}{received signal indicator}
    \acro{SRS}{sounding reference signal}
    \acro{SSB}{synchronization signal block}
    \acro{SINR}{signal-to-interference-plus-noise ratio}
    \acro{SO}{southern}
    \acro{SSS}{secondary synchronization signal}
    \acro{SS}{synchronization signal}
    \acro{SVD}{single value decomposition}
    \acro{thp}{throughput}
    \acro{TRX}{transceiver}
    \acro{UAM}{urban air mobility}
    \acro{UAV}{unmanned aerial vehicle}
    \acro{UE}{user equipment}
    \acro{UL}{uplink}
    \acro{UMa}{urban macro}
    \acro{UMi}{urban micro}
    \acro{UPA}{uniform planar array}
    \acro{UPi}{urban pico}
    \acro{URD}{Urban Random Distributed}
    \acro{URLLC}{ultra-reliable low latency communication}
    \acro{WRC}{world radiocommunication conference}
    \acro{ZF}{zero forcing}
    \acro{ARFCN}{absolute radio frequency channel number}
    \acro{E-ARFCN}{E-UTRA absolute radio frequency channel number}
    \acro{NR-ARFCN}{NR absolute radio frequency channel number}
    \acro{PCI}{physical cell identity}
    \acro{MCC}{mobile country code} 
    \acro{MNC}{mobile network code} 
    \acro{RandS}[R\&S]{Rohde \& Schwarz}
    \acro{RTT}{Round Trip Time}
    \acro{NTN}{Non-Terrestrial Network }
    \acro{UPV}{Universidad politecnica de Valencia}
    
\end{acronym}

\bibliographystyle{IEEEtran}
\bibliography{journalAbbreviations, bibl}

\end{document}